\documentclass[prb,aps,twocolumn,superscriptaddress,showpacs,floatfix]{revtex4-1}

\usepackage[colorlinks=true,linkcolor=black, citecolor=blue, urlcolor=blue, 
    unicode=true]{hyperref}
\usepackage{soul}
\usepackage{color}
\usepackage{graphicx}
\usepackage{amsmath}
\usepackage{blindtext}
\usepackage{gensymb}
\usepackage{bm}
\usepackage[version=4]{mhchem}

\begin{document}
\title{Phase diagram of \ce{CeSb2} from magnetostriction and magnetization measurements: Evidence for ferrimagnetic and antiferromagnetic states}
\author{Christopher Trainer}
\affiliation{SUPA, School of Physics and Astronomy, University of St Andrews, North Haugh, St Andrews, Fife, KY16 9SS, United Kingdom}
\author{Caiden Abel}
\affiliation{Department of Physics and Astronomy and Ames Laboratory, Iowa State University, Ames, Iowa 50011, USA}
\author{Sergey L. Bud'ko}
\affiliation{Department of Physics and Astronomy and Ames Laboratory, Iowa State University, Ames, Iowa 50011, USA}
\author{Paul C. Canfield}
\affiliation{Department of Physics and Astronomy and Ames Laboratory, Iowa State University, Ames, Iowa 50011, USA}
\author{Peter Wahl}
\affiliation{SUPA, School of Physics and Astronomy, University of St Andrews, North Haugh, St Andrews, Fife, KY16 9SS, United Kingdom}

\date{\today}

\begin{abstract} 
Cerium diantimonide (\ce{CeSb2}) is one of a family of rare earth based magnetic materials that exhibit metamagnetism, enabling control of the magnetic ground state through an applied magnetic field. At low temperatures, \ce{CeSb2} hosts a rich phase diagram with multiple magnetically ordered phases for many of which the order parameter is only poorly understood. In this paper, we report a study of its metamagnetic properties by Scanning Tunneling Microscopy (STM) and magnetization measurements. We use STM measurements to characterize the sample magnetostriction with sub-picometer resolution from magnetic field and temperature sweeps. This allows us to directly assess the bulk phase diagram as a function of field and temperature and relate spectroscopic features from tunneling spectroscopy to bulk phases. Our magnetostriction and magnetisation measurements indicate that the low temperature ground state at zero field is ferrimagnetic. Quasiparticle interference mapping shows evidence for a reconstruction of the electronic structure close to the Fermi energy upon entering the magnetically ordered phase. 
\end{abstract}

\maketitle

\section{Introduction}
In a metamagnetic material, the balance between different magnetic interactions means that the order parameter of the ground state sensitively depends on the applied magnetic field. This provides an opportunity to control the ground state, with potential technological implications for spintronics applications. Yet, for many of these materials, the magnetic order parameter and the dominant interactions driving the metamagnetic behaviour remain unresolved. Metamagnetic materials can broadly be subdivided into two types, depending on whether the phase transition is first or second order. If the transition is second order and can be continuously tuned to zero Kelvin a quantum critical point occurs, and the transition at very low temperatures is driven by quantum fluctuations\cite{Gegenwart2008}. An example of a first order metamagnetic transition is the ``spin flop'' transition in an antiferromagnet \cite{Stryjewski}. The change of the order parameter, in both cases, is accompanied by a reconstruction of the Fermi surface or by a Lifshitz transition \cite{Keimer,Lifshitz}. The different natures of the two types of transition are determined by what drives the transition. For first order metamagnetic transitions, the magnetic moments are large and localized in nature and transitions in magnetic order occur due to competition between the Zeeman energy and free energy of the different magnetic states. 
This type of metamagnetic system can, if coupled with a strong magneto-crystalline anisotropy, produce a complex magnetic phase diagram. Examples of this class of materials include the rare earth compounds \ce{DyAgSb2} \cite{PhysRevB.59.1121}, \ce{HoNi2B2C}\cite{CanfieldHo}, \ce{CeSb} \cite{WIENER2000505}, \ce{TbPtIn} and \ce{TmAgGe} \cite{TbPtIn}.
In the case of the second order quantum critical metamagnetic transition, electronic correlations play an important role to stabilize the magnetic order and the metamagnetic phase diagram is often dominated by a competition between Kondo physics and the RKKY interaction \cite{doniach_kondo_1977,Gegenwart2008,MISRA2008vii}. 

\begin{figure}[bt!]
\begin{center}
\includegraphics{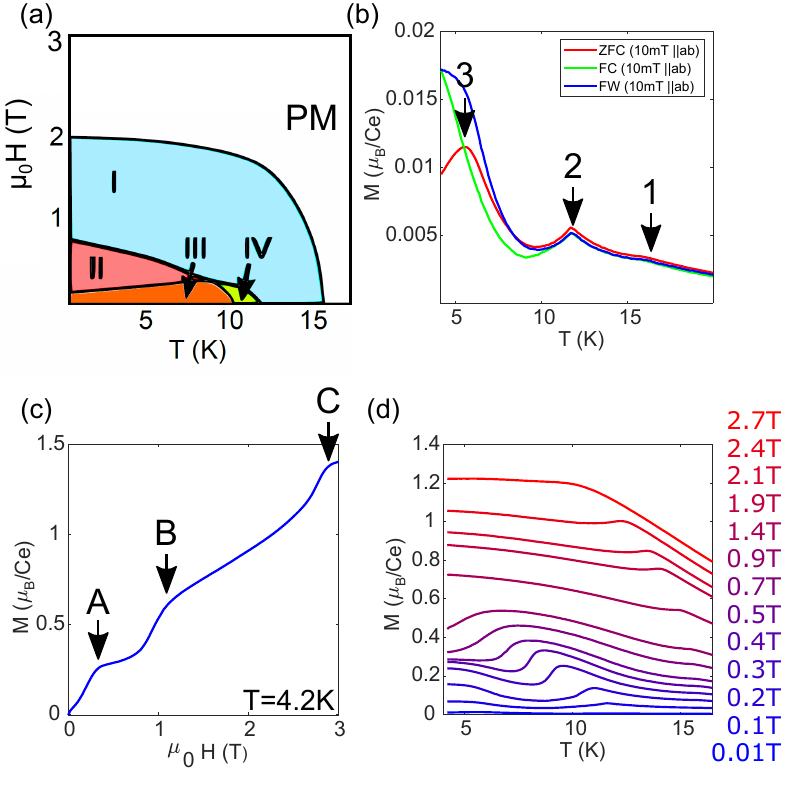}
\end{center}
\caption{\textbf{Magnetic measurements of \ce{CeSb2}} (a) Phase diagram of \ce{CeSb2} following ref.~\onlinecite{BudKo1998}. Four different magnetic phases have been identified (labelled I-IV) in addition to the paramagnetic phase (PM) at high temperatures $T>15.5\mathrm{K}$. (b) Zero field cooled (ZFC), field cooled (FC) and field warmed (FW) magnetization measurements all recorded using an applied field of $10\mathrm{mT}$. Three transitions can be observed at 1, 2 and 3. (c) Magnetic isotherm recorded at $T=4.2\mathrm{K}$ for fields up to $3\mathrm{T}$. The three kinks at $0.4\mathrm{T}$ (A), $1\mathrm{T}$ (B), and $2.8\mathrm{T}$ (C) in the isotherm are associated with metamagnetic transitions. (d) Zero field cooled magnetization curves as a function of temperature and for magnetic fields up to $2.7\mathrm{T}$.}
\label{figure_1_new}
\end{figure}

Here, we study \ce{CeSb2}, a rare-earth metamagnetic material which undergoes multiple phase transitions as a function of field and temperature \cite{BudKo1998}. \ce{CeSb2} shows signatures of Kondo-lattice-like behaviour in resistivity measurements below room temperature \cite{BudKo1998,Zhang_2017} and evolves through a number of magnetic phase transitions as the temperature is lowered. A sketch of the phase diagram is shown in fig.~\ref{figure_1_new}(a). The first magnetic phase transition from the paramagnetic (PM) phase occurs at $15.5\mathrm{K}$ to a phase I and can be seen in resistivity \cite{BudKo1998,Zhang_2017}, magnetization \cite{BudKo1998,Zhang_2017} and specific heat measurements \cite{SuderowCDW}. Further magnetic transitions to phases IV and III are observed at $\sim 11.5\mathrm{K}$ and $9.5\mathrm{K}$ \cite{BudKo1998,Zhang_2017,SuderowCDW,liu_neutron_2020}. The nature of the ground state at zero field, phase III, is controversial: Magnetization measurements  show a large magnetic response \cite{BudKo1998,Zhang_2017} indicating a ferromagnetic ground state however neutron scattering shows the distinct onset of a magnetic ordering vector $(\frac{1}{6},1,0)$ at $9.8\mathrm{K}$ \cite{liu_neutron_2020}. This ordering vector persists to lower temperatures, pointing towards an antiferromagnetically ordered phase. With increasing field from phase III, the system exhibits a transition into an intermediate phase II before entering phase I and at even higher fields the paramagnetic phase. In this paper, we use low temperature scanning tunneling microscopy and magnetization measurements to study the low temperature phases of \ce{CeSb2} and elucidate upon the nature of the different magnetically ordered states and the accompanying changes in the electronic structure. 

\section{Experimental Methods}
\subsection{Sample growth}
Crystals were grown as described in Refs.~\onlinecite{Canfield1991,crystalgrowth2}. Large ($15\times15\times1 \mathrm{mm}^3$) single crystals of \ce{CeSb_2} were grown out of antimony flux.\cite{Canfield1991,crystalgrowth2} Elemental Ce and Sb, in a 7:93 atomic ratio, were placed in a fritted crucible set,\cite{canfield_use_2016} sealed in a silica tube and cooled from $1175^\circ\mathrm{C}$ to $750^\circ\mathrm{C}$ at which point the excess Sb was decanted with the aid of a centrifuge.\cite{canfield_new_2020} The crystals
grow as soft plates with the c-axis perpendicular to the
plates. The samples are micaceous, and layers can be easily
peeled off. Each of these layers is malleable but, despite this
ease of deformation, very high residual resistance ratios are found \cite{BudKo1998}.

\subsection{Scanning tunneling microscopy}
We use a home-built low temperature STM mounted in a vector magnet capable of applying a magnetic field of up to $5\mathrm{T}$ in any direction relative to the sample and at temperatures to a base temperature of 2K \cite{Trainer2017}. Samples were cleaved in-situ in cryogenic vacuum at a temperature of about $20\mathrm{K}$ and then directly inserted into the STM head to prepare atomically flat and clean surfaces.

\subsection{SQUID magnetization measurements}

Magnetization measurements were carried out in a Quantum Design MPMS. The sample was aligned such that the $c$ axis was perpendicular to the applied field.  The bulk of the data, unless otherwise stated,  was measured whilst warming the sample after having cooled in zero applied field.

\section{Results}
\subsection{Magnetic susceptibility}
 \ce{CeSb2} exhibits a complex phase diagram at low temperature (figure \ref{figure_1_new}a).\cite{BudKo1998} Attempts have been made to characterize the low temperature magnetic phases by multiple groups \cite{BudKo1998,Zhang_2017,liu_neutron_2020} each resulting in slightly different magnetic phase diagrams. As the temperature is lowered in zero field, \ce{CeSb_2} undergoes a first magnetic transition at $15.5\mathrm{K}$ to magnetic state I. Further cooling causes \ce{CeSb_2} to transition to a second magnetic phase (IV) at around $12\mathrm{K}$ before finally settling to its ground state (III) below $10\mathrm{K}$. Once \ce{CeSb_2} has reached its low temperature ground state it can be tuned through multiple magnetic states (II and I) by the application of a magnetic field in the sample's $ab$ plane. The exact nature of the magnetic order in each of these states (I-IV) is still unclear.          
 
 We have conducted magnetization measurements with a field applied in the sample $ab$ plane in order to better understand the nature of these bulk magnetic phases of \ce{CeSb_2}. Figure~\ref{figure_1_new}b shows temperature dependent measurements of the sample magnetization. From the zero field cooled measurements of the sample (red curve in figure \ref{figure_1_new}b) we detect three transitions. When cooling down, the first transition (1) which corresponds to the onset of magnetic phase I is found at $15.6\mathrm{K}$ where the magnetization undergoes a slight change of slope. The second transition (2) is observed as a sharp downturn in the magnetization at $11.4\mathrm{K}$. This downturn of the magnetic susceptibility is indicative of a transition to an antiferromagnetic state \cite{Kittel2004,BUBE1992242}. A final transition (3) is seen at $6.5\mathrm{K}$ where the sample's zero field cooled magnetization exhibits a maximal value. Field cooling (FC) and warming (FW) measurements show a substantially larger low temperature magnetization compared to the zero field cooled measurement indicates  the formation of either ferromagnetic or ferrimagnetic domains in these phases \cite{Joy_1998}. 
 
 To map out the full $H$-$T$ phase diagram, we have recorded the magnetisation in isothermal field sweeps. Figure \ref{figure_1_new}c shows one such measurement at $4.2\mathrm{K}$. Three prominent kinks (A,B,C) are observed. Analyzing the peaks in the derivative $\frac{\partial M}{\partial H}$, we identify that the three field-induced transitions occur at $0.2\mathrm{T}$ (A), $0.9\mathrm{T}$ (B) and $2.8\mathrm{T}$ (C). These transitions are in broad agreement with those reported previously \cite{BudKo1998,Zhang_2017,liu_neutron_2020}, with transition C seen at slightly higher fields. 
 
 We have further conducted zero field cooled magnetization measurements for a range of applied fields, Figure~\ref{figure_1_new}d. As the applied field is increased, transition 2 occurs at a decreasing temperature before merging with transition 3. Further increase of field suppresses both transitions to lower temperatures. Transition 1 can be seen to follow an almost mean-field-like behaviour as a function of applied field and temperature. 
 
\begin{figure}[t]
\includegraphics{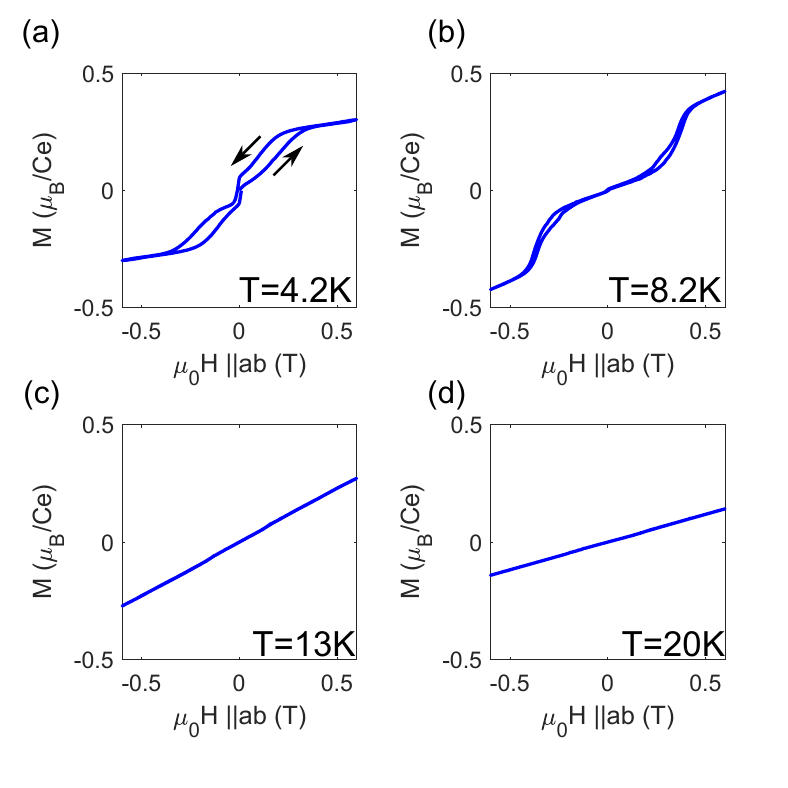}
\caption{\textbf{$M$-$H$ loops.} (a) $M$-$H$ loop recorded at $4.2\mathrm{K}$ in the lowest temperature magnetic phase (phase III in fig.~\ref{figure_1_new}). Clear hysteretic behaviour can be seen. (b) $M$-$H$ loop recorded at $8.2\mathrm{K}$. The sample shows less hysteresis in this phase and a linear magnetic response at low field. (c) $M$-$H$ loop recorded in phase I, showing only a paramagnetic response. (d) $M$-$H$ loop recorded in the paramagnetic phase at $T=20\mathrm{K}$. The sample shows a smaller coupling to the field than in phase I. Between the measurements of $M$-$H$ loops at different temperatures, the sample was warmed up to $20\mathrm{K}$ and then cooled in zero field.}
\label{s3}
\end{figure}

The nature of the low temperature ground state has been the object of speculation with both ferromagnetism and antiferromagnetism being proposed.\cite{Canfield1991,BudKo1998,Zhang_2017,liu_neutron_2020} To shed new light on the nature of the ground state, we have measured the magnetization as a function of applied field in $M$-$H$ loops with field in the $ab$-plane. Magnetization measurements in the low temperature phase (Figure~\ref{s3}a) reveal hysteretic behaviour with a strong reduction of the magnetization near zero applied field, similar to what is observed in ferrimagnetic materials.
The interpretation in terms of ferrimagnetism in the low temperature phase is consistent with neutron scattering measurements \cite{liu_neutron_2020} which detect magnetic order below $9.8\mathrm{K}$. Magnetization measurements of the intermediate temperature phase show a linear magnetic response at low fields (Fig.~\ref{s3}b). This fact, coupled with the suppressed magnetic susceptibility in this phase and AFM-like phase transition lead us to conclude that this phase is antiferromagnetic. Measurements in phase I, figure~\ref{s3}c, and at temperatures above the magnetic transitions, figure~\ref{s3}d, show a linear response. When crossing the transition from phase I to the paramagnetic phase with increasing temperature from 13K, the magnetic susceptibility reduces to about half that in phase I at the phase transition.

\begin{figure*}[t]
\includegraphics[width=170mm]{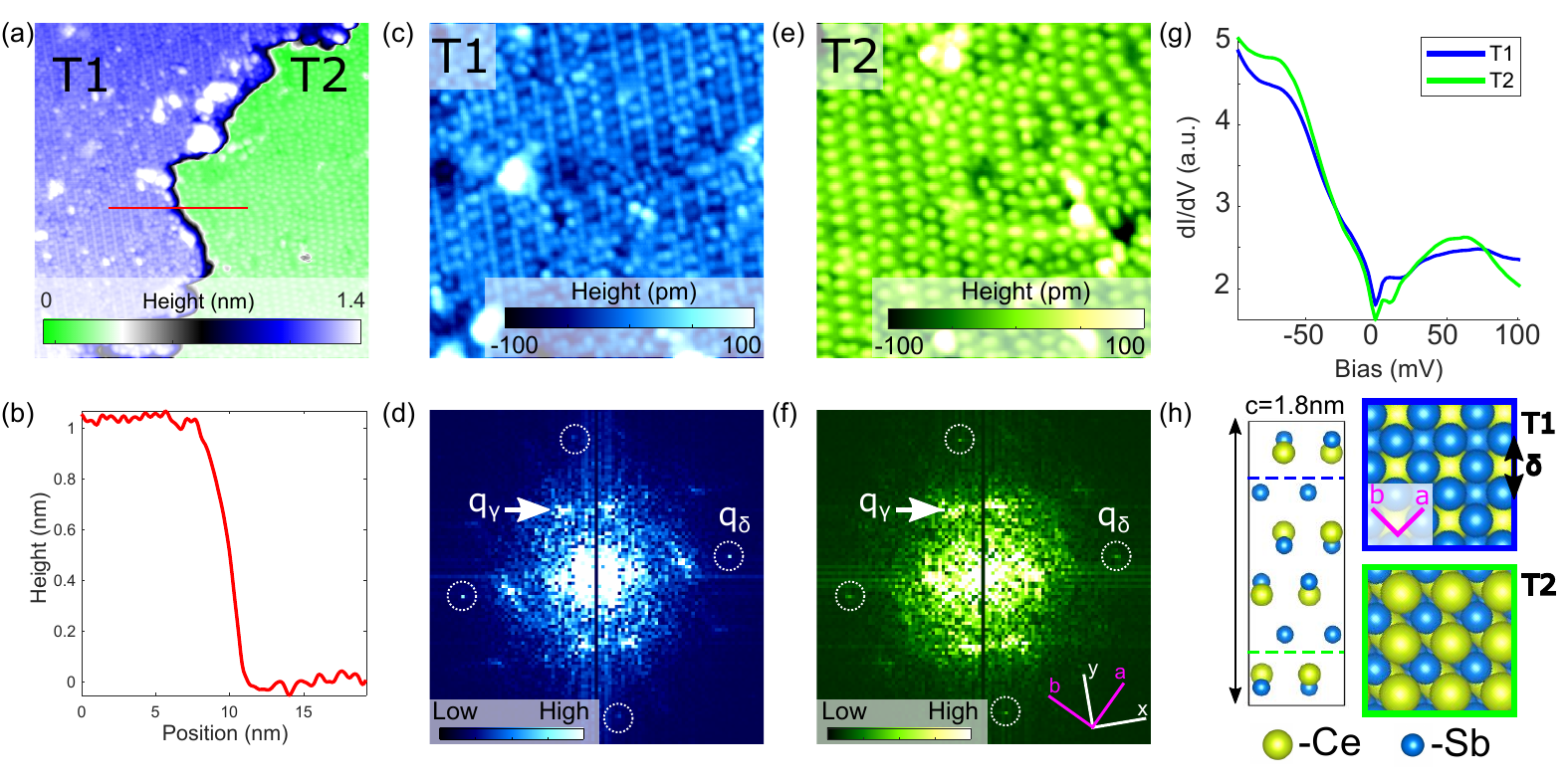}
\caption{\textbf{Topographic imaging of \ce{CeSb2}.} (a)  Topographic STM image of a step edge showing two different surface terminations T1 and T2 of \ce{CeSb2} ($V=100\mathrm{mV}$, $I=40\mathrm{pA}$, $(50\times30)\mathrm{nm}^2$, $T=2.2\mathrm{K}$) (b) The line profile taken along the red line in (a). (c) Topographic image showing the termination labelled T1 attributed to the Sb layer ($V=100\mathrm{mV}$, $I=40\mathrm{pA}$, $(20\times20)\mathrm{nm}^2$, $T=2.2\mathrm{K}$).  (d) The Fourier transformation of the topographic image shown in (c). (e) Topographic STM image of the termination labelled T2 attributed to the CeSb layer ($V=100\mathrm{mV}$, $I=40\mathrm{pA}$, $(20\times20)\mathrm{nm}^2$, $T=2.2\mathrm{K}$). (f) The Fourier transformation of (e). (g) Average STS spectra recorded on the different terminations taken from a spectroscopic map across the step in (a) ($V_\mathrm{s}=100\mathrm{mV}$, $I_\mathrm{s}=100\mathrm{pA}$, $V_{\mathrm{mod}}=2.5\mathrm{mV}$). (h) Crystal structure of \ce{CeSb}. The Sb termination (tentatively attributed to T1) is indicated by the blue dashed line, the blue box on the right shows its surface structure. The CeSb termination (T2) is indicated by the green dashed line with its surface structure in the green box.
}
\label{figure1}
\end{figure*}

\subsection{STM Topographic imaging and basic spectroscopy}
Larger scale topographic STM images of the surface of \ce{CeSb2} acquired at a temperature of $2.2\mathrm K$ show atomically flat terraces with occasional step edges (see Fig.~\ref{figure1}a). The terraces exhibit two predominant surface structures, representing two different surface terminations. Measurements of steps between the terminations (Fig.~\ref{figure1}b) yield a step height of $1\mathrm{nm}$, less than the unit cell height ($1.8\mathrm{nm}$),  broadly consistent with the expected fractional step height of $1.2\mathrm{nm}$ between different terminations. The crystal structure suggests that the two possible terminations are an Sb-terminated surface and a \ce{CeSb}-terminated surface (Fig.~\ref{figure1}h). STM images focusing on the two terminations reveal a complex surface structure (Fig.~\ref{figure1}c, e). We label the two surface terminations T1 (Fig.~\ref{figure1}c) and T2 (Fig.~\ref{figure1}e). Images of both terminations show an atomic lattice with almost tetragonal symmetry and a periodicity of $4.7\mathrm\AA$, which can be clearly seen in the Fourier transformation of the images (labelled $\mathbf{q}_\delta$ in Figure~\ref{figure1}d and f). For the CeSb termination, this periodicity is in good agreement with the atomic distances between \ce{Ce} atoms expected at the surface, whereas for the Sb-terminated surface, this suggests an additional superstructure resulting in an inequivalence of the surface Sb atoms (Fig.~\ref{figure1}h). Based on this, we identify the $q_\delta$ direction as the crystallographic [110] axis. Both terminations exhibit complex superstructures superimposed on the atomic lattice. These structures manifest as $C_2$ symmetric ladder-like patterns in the case of T1 and  bubble-like patterns in the case of T2. Both superstructures give three distinct q vectors arranged in a quasi hexagonal-like structure in the Fourier transformation similar to the patterns found in La$_x$Ce$_{1-x}$Sb$_2$\cite{SuderowCDW}. Both show peaks at $q_\gamma=(\pm\frac{1}{12},\frac{1}{2})$ (relative to the atomic peaks at $\mathbf{q}_\delta$) and peaks at $q=(\frac{2}{3},0)$ for T1 and $q=(\frac{1}{3},0)$ for T2 . They remain stable up to temperatures above the magnetic phase transitions (see appendix, figure~\ref{s1}). Bias dependent images show phase reversal of the ``bubble" patterns (see appendix, figure~\ref{s10}) providing evidence that at the surface of \ce{CeSb2} we observe a charge density wave (CDW) again similar to the surface of La$_x$Ce$_{1-x}$Sb$_2$ \cite{SuderowCDW}. 
We tentatively attribute T1 as an Sb terminated surface and T2 as CeSb termination (see appendix \ref{appendix-t1}). 

Spectroscopic characterisation of the two surface terminations (figure~\ref{figure1}g) shows that they exhibit broadly similar features with a slight suppression of the differential conductance at and just above the Fermi level for termination T2 in comparison to T1.

We have conducted spectroscopic mapping on both terminations, however, very low quasi particle scattering interference signal was observed on T1 so in the following we will focus our discussion on the quasi-particle interference for termination T2.  

\subsection{Quasi-particle Interference and field and temperature dependent spectroscopy}
\begin{figure}[bt!]
\begin{center}
\includegraphics{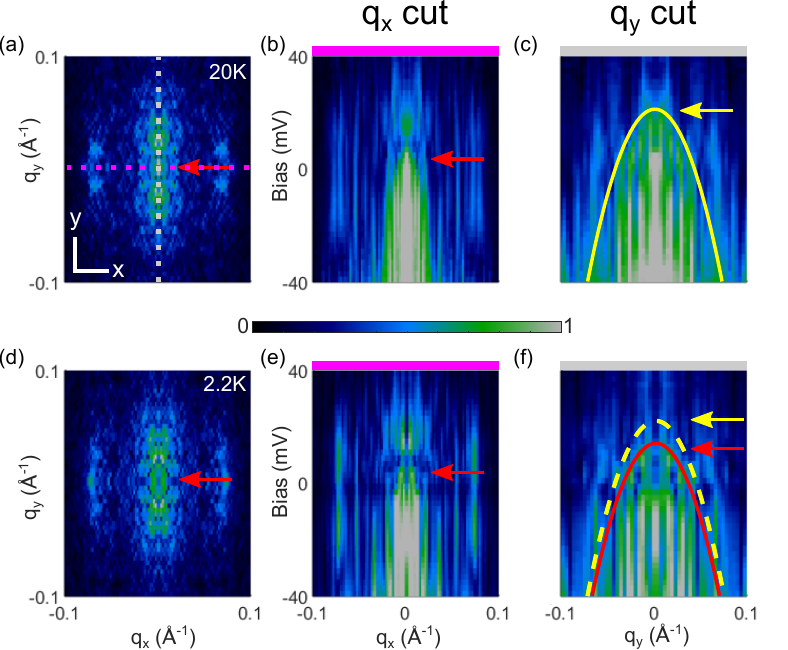}
\end{center}
\caption{ \textbf{Temperature-dependent Quasiparticle Interference.} (a) Fourier transformation (symmetrized) of a map layer at a bias voltage of $+6\mathrm{mV}$ recorded in the high temperature paramagnetic phase ($T=20\mathrm{K}$).
(b, c) Linecuts through the FT-STS data shown in (a) along the $x$- and $y$-directions marked in (a). Red arrows highlight the presence of scattering between bands. (d) Fourier transformation (symmetrized) of an STS map layer with the same bias voltage as in (a) recorded in the low temperature phase ($T=2.2\mathrm K$). (e,f) Linecut through the FT-STS data along the $x$- and $y$-directions marked in (a). Red arrows in $q_x$ data (b,e) highlight the absence of the scattering that was present at $20\mathrm{K}$. (c) Yellow parabola is fit to $20\mathrm{K}$ data with maximum at $22.5\mathrm{mV}$. In the $2.2\mathrm{K}$ data (f) the red solid line is a fit to the data with maximum at $14\mathrm{mV}$, the $20\mathrm{K}$ fit is shown as a dashed yellow line.  All spectroscopic data shown in (a)-(f) were recorded in the same atomic location and with the same tip and identical tunnelling conditions ($\mathrm{V}_{\mathrm{s}}=100\mathrm{mV}$, $\mathrm{I}_{\mathrm{s}}=1\mathrm{nA}$,  $\mathrm{V}_{\mathrm{mod}}=2\mathrm{mV}$).}  
\label{figure4}
\end{figure}
Establishing how the low energy electronic structure changes across the metamagnetic phase transitions can provide key insights for a microscopic model to understand the phase diagram. Here, we use spectroscopic maps to study how the electronic structure near the Fermi level changes between the high temperature paramagnetic state and the low temperature phase. We have recorded these spectroscopic maps on termination T2, the one we tentatively assign to a \ce{CeSb} termination, at temperatures above ($T=20\mathrm{K}$) and below the phase transition ($T=2.2\mathrm{K})$. These maps were measured in the same atomic-scale location on the sample, with the exact same STM tip and identical tunneling conditions. In the Fourier transformation of the spectroscopic data we can distinguish several scattering vectors. In fig.~\ref{figure4}a and d we show the Fourier transformations of these maps at $20\mathrm{K}$ and $2.2\mathrm{K}$, respectively. Linecuts taken along the $x$ direction through the data obtained in the paramagnetic phase are shown in figure \ref{figure4}b and show a weakly dispersing $\mathbf{q}$-vector at approximately $0.02\mathrm\AA^{-1}$. Similar linecuts from the data recorded in the low temperature phase (fig.~\ref{figure4}e) show that the states just above the Fermi level (between $+2\mathrm{mV}$ and $+10\mathrm{mV}$) appear to increase their mass and a gap appears in the scattering pattern. 

Similar linecuts taken along the $y$ direction are shown in figure~\ref{figure4}c and f. In the paramagnetic phase at $20\mathrm{K}$ (figure~\ref{figure4}c) we can identify a hole-like dispersion with a band maximum at $22.5\pm2\mathrm{mV}$. The same cut taken in the low temperature phase shows that this band maximum has shifted quite substantially towards the Fermi level with the maximum now at $14\pm1\mathrm{mV}$. This shift in the energy of the band edge could be indicative of magnetic exchange splitting in the low temperature phase similar to that found in \ce{CeSb}\cite{CeSbexchangesplit}.  
\begin{figure}[bt!]
\begin{center}
\includegraphics{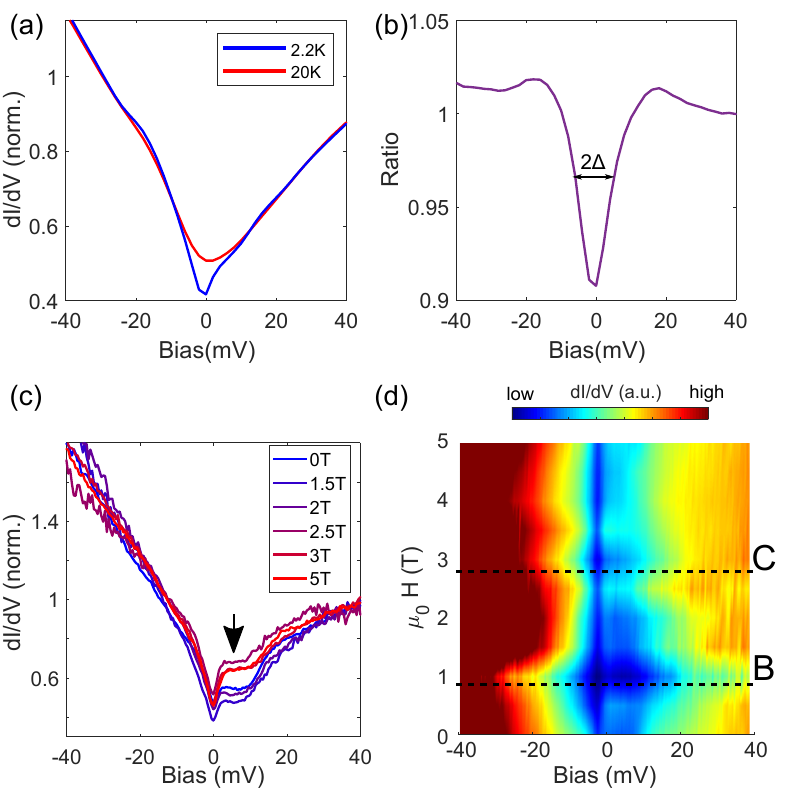}
\end{center}
\caption{ \textbf{Temperature and field dependent spectroscopy.} (a) Averaged dI/dV spectroscopy recorded with the Quasiparticle interference maps discussed in figure~\ref{figure4}. Spectra were recorded with the same tip and at the same location on the sample surface ($\mathrm{V}_{\mathrm{s}}=100\mathrm{mV}$, $\mathrm{I}_{\mathrm{s}}=1\mathrm{nA}$,  $\mathrm{V}_{\mathrm{mod}}=2\mathrm{mV}$).(b) The ratio of the spectrum taken at $2.2\mathrm{K}$ thermally broadened to $20\mathrm{K}$ and the spectrum acquired at $20\mathrm{K}$. A gap can be seen around the Fermi level which has a half width of $\Delta=6\mathrm{mV}$. (c) Field dependent point spectra recorded with the same tip apex and averaged over multiple points on the sample surface. The Field was applied in the sample $ab$ plane $16\degree$ from $q_\delta$. The sample density of states can be seen to increase in a $10\mathrm{mV}$ window above the Fermi level for fields above $2T$ indicated by the arrow. (d) Color plot of the tunneling spectra shown in c as a function of applied field.} 
\label{figure_spectra}
\end{figure}

Looking at the spatially averaged differential conductance spectra (figure~\ref{figure_spectra}a and b), we find that a $\sim 6\mathrm{mV}$ gap opens around the Fermi level when the sample is cooled to the lowest temperature phase. The substantial reconstruction of the scattering pattern can have multiple origins, the most likely of which are: (1) the onset of a spin density wave state or antiferromagnetic order that leads to a reconstruction of the Fermi surface and the opening of hybridization gaps; (2) opening of a hybridization gap due to interaction of localized states, e.g. $f$ electrons, with delocalized states\cite{Schmidt2010, Aynajian2012}.

To examine how the sample's electronic structure is affected with the application of a field in the $ab$ plane we have recorded differential conductance spectra as a function of field (figure~\ref{figure_spectra}c and d, field applied at $16^{\circ}$ to the $y$ direction). The spectra were recorded with the same tip apex and averaged over multiple points on the T2 terminated surface. The overall line shape of the differential conductance spectra does not change with applied field as can be seen in figure~\ref{figure_spectra}c except in a $10\mathrm{mV}$ window above the Fermi level. We observe that for applied fields of $2\mathrm{T}$ and lower in the $ab$ plane we see a suppressed density of states in this energy window. For fields of $2.5\mathrm{T}$ and above, the density of states in this window increases. This potentially indicates that the surface has gone through a magnetic transition at $\sim 2\mathrm{T}$ and that the bands in this energy range of $10\mathrm{mV}$ above the Fermi level strongly couple to the field. In our bulk measurements of \ce{CeSb_2} we have not observed a transition between $2$ and $2.5\mathrm{T}$ indicating that the magnetism at the surface could be different to that of the bulk. Figure~\ref{figure_spectra}d shows how the spectroscopy evolves as a function of field in comparison to the transitions in the bulk which are highlighted for reference.

\subsection{Magnetostriction}

\begin{figure*}[t]
\includegraphics[width=170mm]{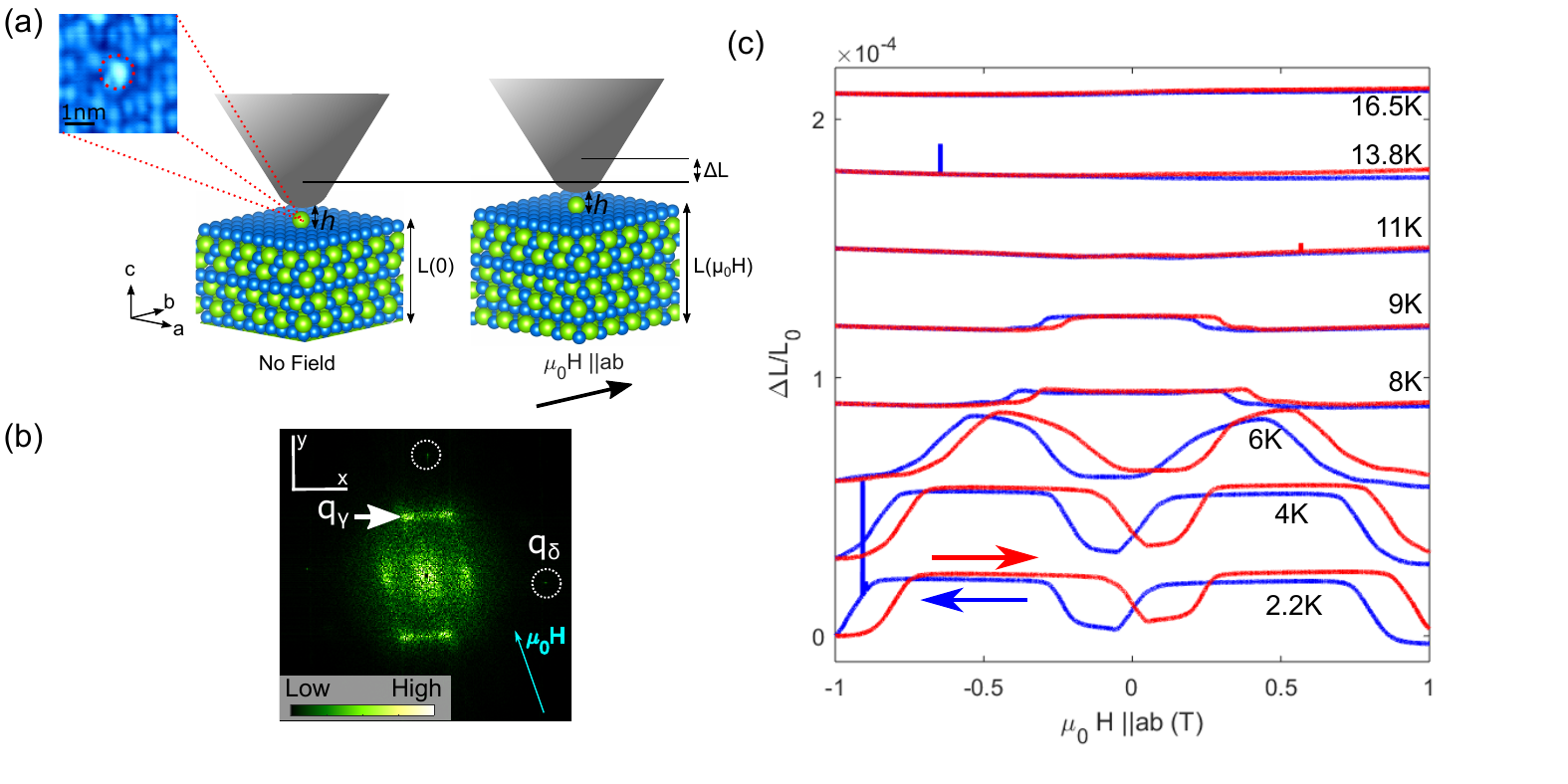}
\caption{\textbf{Magnetostriction measurements in STM.} (a) Schematic of the principle of how an STM can be used to measure the magnetostriction of a sample. The STM tip is held at a constant current $I$, and thus a constant height above the surface, over a defect (shown in the inset) while the magnetic field is ramped. To keep the tip in the same lateral position, an atom tracking algorithm is used. The magnetic-field-induced change $\Delta L$ in the sample $c$ axis leads to a change in the tip position to keep the tunneling current constant, allowing for a measurement of the magnetostriction. (b) Fourier transformation of a topographic image of the \ce{CeSb} termination (T2). The cyan arrow indicates the direction of the applied magnetic field $16\degree$ from the [110] direction. (c) Measurement of magnetostriction of \ce{CeSb2} in the low temperature phase at fixed temperature as a function of applied field. Clear magnetic hysteresis behaviour can be seen between the forward and backward sweeps. The sweeps are offset from each other by $3\times10^{-5}$ and $0$ is defined at a field of $-1\mathrm{T}$. Data between $2$ and $6\mathrm{K}$ are in the low temperature phase (phase III). Data between $8$ and $11\mathrm{K}$ are in the intermediate temperature phase (phase IV). The $13.8\mathrm{K}$ data is in phase I and the $16.5\mathrm{K}$ data is outside the magnetically ordered phases. 
}
\label{figure2}
\end{figure*}
To be able to correlate surface properties with the bulk magnetic phase diagram, we have studied the magnetostriction of \ce{CeSb2}, using the STM as a dilatometer. Measurement of the thermal expansion in STM has been demonstrated previously \cite{crespo_thermal_2006,galvis_magnetic_2012}, here we complementarily also measure the tip height and thus sample thickness during magnetic field ramps. Rare earth compounds often display comparatively large magnetostriction \cite{Liu2012,KOON1971413,Clark}. To measure the expansion/contraction we track the vertical position of the STM tip while ramping the magnetic field or the temperature (compare Figure~\ref{figure2}a). During this measurement, the tip-sample distance is kept constant by maintaining a constant tunneling current. In order to separate the expansion in the direction of the surface normal from magnetostriction effects in the lateral direction, the tip position is locked onto a surface defect throughout the measurement by an atom tracking algorithm \cite{nanonis-atom}. The algorithm adjusts the tip position such that in the tip's reference frame the defect is static. When sweeping the applied magnetic field or temperature and recording the change in the STM tip's vertical position, we thus measure only the expansion and contraction of the sample along its $c$ axis and can extract the change in sample thickness as a function of temperature $\Delta L(T)$ or magnetic field $\Delta L(H)$.

Figure~\ref{figure2}c shows $\Delta L(H)$ curves obtained at a temperature of $4\mathrm{K}$ and with a field $H$ applied in a direction $16\degree$ from the [110] crystallographic axis (indicated in Figure~\ref{figure2}b). The physical dimensions of the sample show a strong response to the applied magnetic field with size of the sample in the $c$ direction expanding by up to $6\mathrm{nm}$ over the total sample thickness $L_0=440\mu\mathrm{m}$. This translates to a $0.024\mathrm{pm}$ expansion per CeSb$_2$ unit cell. This level of response is typical for rare earth ferromagnetic materials \cite{Liu2012} and is also a feature observed in the magnetostrictive response of ferrimagnetic materials \cite{Sukhorukov2018}. The behaviour of the magnetostriction is difficult to understand in terms of an antiferromagnetic low temperature phase where the magnetostriction should only exhibit a small response around zero field \cite{Doerr}. 
We measured magnetostriction curves at a set of temperatures to explore the magnetic phase diagram. Figure~\ref{figure2}c shows a $\Delta L(H)$ curve measured at a temperature of $8\mathrm{K}$, pushing the sample into its second magnetic phase between $6.5\mathrm{K}$ and $11.4\mathrm{K}$. The response of the sample to the applied field is dramatically changed. For small fields close to zero there is practically no change in the length along the $c$ axis showing that the sample has entered into a different magnetic phase. The shape of the magnetostriction curve at low field is closer to what one would expect for an antiferromagnet.

\begin{figure*}
\includegraphics[width=150mm]{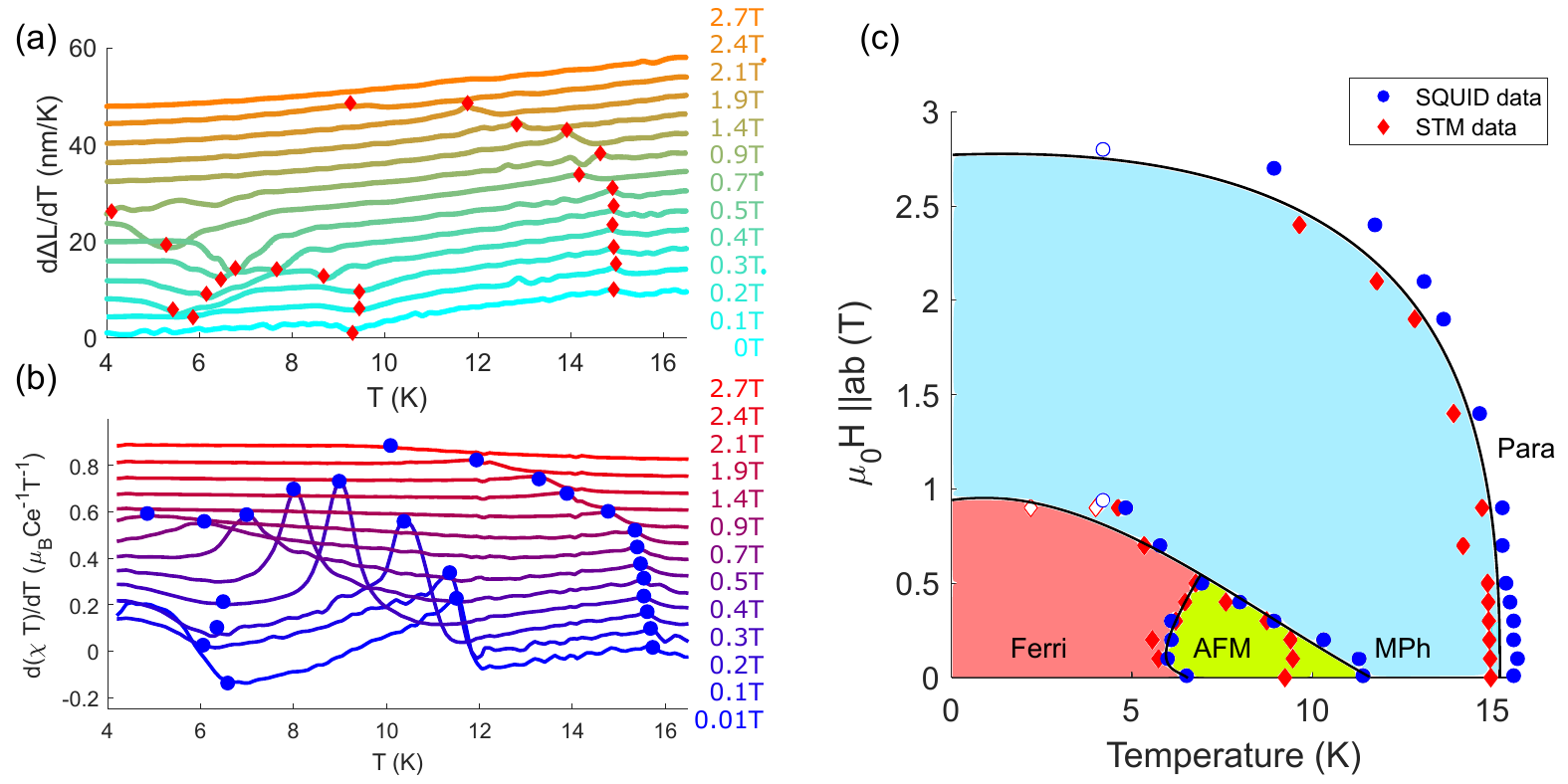}
\caption{\textbf{Magnetic phase diagram of \ce{CeSb2}.} (a) Temperature derivative of the thermal contraction, $\frac{d\Delta L(T)}{dT}$, measured in the STM at different magnetic fields $\mu_0 H$ applied in the $ab$ plane. Between measurements, the sample was warmed up to $20\mathrm{K}$ and then cooled in zero field. (for raw data see appendix, figure~\ref{s6}). 
(b) Temperature derivatives of the magnetic susceptibility times temperature, $\frac{d (\chi(T)T)}{dT}$. Curves in (a) and (b) are vertically offset for clarity. (c) Revised phase diagram of \ce{CeSb2} as a function of temperature and magnetic field constructed from magnetostriction data shown in (a) and magnetization in (b). Filled data points were obtained from warming sweeps recorded at constant field. Open data points were recorded from isothermal measurements with increasing field. Ferri correspond to a ferrimagnetic phases, AFM is an antiferromagnetic phase, MPh a magnetic phase with unknown order and Para the high temperature paramagnetic phase.
}
\label{figure3}
\end{figure*}

Figure \ref{figure2}c shows magnetostriction curves measured using the STM throughout the entire low temperature phase diagram of \ce{CeSb2} for fields up to $1\mathrm{T}$. For temperatures between $2$ and $6\mathrm{K}$ a pronounced response is seen in the magnetostriction, showing an increase in the $c$-axis height until about $0.2\mathrm{T}$, followed by a sharp decrease at $\sim0.9\mathrm{T}$. The plateau between $\sim0.2\mathrm{T}$ and $\sim 0.8\mathrm{T}$ decreases in width with increasing temperature up to $6\mathrm K$. Between $6\mathrm K$ and $8\mathrm{K}$, the magnetostrictive behaviour of the sample changes drastically. At low fields, there is very little change in the sample's $c$ axis as a function of field, a response that is indicative of an antiferromagnet \cite{magnetostrictionAFM2,Doerr,magnetoAFM1}. As the field is increased, the sample contracts as it crosses into another phase at $\sim 0.4\mathrm{T}$, however significantly less than the change in $c$-axis height seen at lower temperatures. The last vestige of this ``top hat''-like feature in the magnetostriction data is seen at $11\mathrm{K}$. For higher temperatures no evidence of phase transitions is seen in the isothermal sweeps.  

Previous measurements of the phase diagram of \ce{CeSb2} have shown a field induced transition at $\sim0.2T$ at low temperatures. We also observe this transition as a kink (A) in figure \ref{figure_1_new}c. Measuring hysteresis loops, using both magnetostriction (figure \ref{figure2}) and magnetization (figure \ref{s3}), across this field induced transition shows results that are most consistent with ferrimagnetic behaviour. We therefore conclude that this kink (A) in the magnetization arises due to the complete polarization of the low temperature ferrimagnetic state. Hence we conclude that phases II and III (figure \ref{figure_1_new}a) are not independent phases but the same phase. 
\subsection{Magnetic phase diagram}  
We have complemented the magnetostriction measurements by measurements of the thermal contraction, where we have recorded the vertical position of the STM tip, $z$, as a function of temperature for various magnetic fields applied in the $ab$ plane of the sample at the same $16\degree$ angle from [110]. The temperature derivatives of this data are shown in Figure~\ref{figure3}a. Transitions between different magnetic phases can clearly be seen as kinks in the $\Delta L(T)$ curves. By tracking the local maxima or minima in the derivative of the $\Delta L(T)$ curve it is possible to construct a magnetic phase diagram. 
For comparison, we show the phase transitions extracted from the magnetic susceptibility in fig.~\ref{figure3}b. Combining both the magnetization and magnetostriction measurements we are able to construct a revised phase diagram at low temperatures with fields up to $3\mathrm{T}$ applied in the sample $ab$ plane, shown in Figure~\ref{figure3}c. 

Upon cooling, the sample first undergoes a transition from a high temperature paramagnetic state to an antiferromagnetic state at $15.6\mathrm{K}$. Due to the second order nature of this transition and the fact that the phase transition exhibits a strong field dependence and we do not observe hysteresis in M-H curves, this indicates antiferromagnetism, however as no ordering vector has been observed in this phase in neutron scattering measurements\cite{liu_neutron_2020} we can only conclude that it is a magnetic phase (MPh) distinct from the high temperature paramagnetism. The magnetostriction measurements show the onset of this transition at a slightly lower temperature (by about $0.5\mathrm{K}$) than the magnetisation but otherwise good agreement is found between both techniques. The discrepancy may arise from differences in how the transition temperature is determined from the data. Cooling the sample further in zero field, the material enters into a new antiferromagnetic phase, the onset of which is found in magnetization at $11.4\mathrm{K}$. Magnetostriction measurements of the onset of this phase show a broad peak in the derivative with respect to temperature centered at $9.2\mathrm{K}$. Previous reports have detected multiple phase transitions in this temperature window \cite{BudKo1998,Zhang_2017,SuderowCDW}. In particular measurements of the heat capacity show two clearly distinct transitions in this temperature window at $12\mathrm{K} $ and $9\mathrm{K}$ \cite{SuderowCDW} leading to the possibility that the magnetization and magnetostriction measurements are sensitive to different phases. Alternatively, another possible explanation for this could be that the magnetic field was applied at a different angle in the $ab$ plane between the SQUID and STM measurement. Alternatively, the differences may stem from the width of this transition, which appears rather broad in the magnetostriction data (see figure~\ref{figure3}a). Finally, at the lowest temperatures below $6.5\mathrm{K}$, the sample enters a ferrimagnetic phase with a small ferromagnetic component and a sizeable magnetization induced expansion in the $c$ direction as the sample is polarized by a field in the $ab$ plane.

\section{Discussion}
We have determined and confirmed the magnetic phase diagram of \ce{CeSb2} at low temperatures by combining high-precision STM-based magnetostriction with sub-picometer resolution along with magnetization measurements. From the magnetization and magnetostriction data, we were able to find three distinct phase transitions at $15.6\mathrm{K}$, $\sim 10\mathrm{K}$ and $6.5\mathrm{K}$, and therefore determine that between $20\mathrm{K}$ and the lowest temperatures \ce{CeSb2} exhibits four magnetic phases. From measurements of the sample magnetization and magnetostriction as a function of field we characterize these phases as a high temperature paramagnetic phase, a magnetically ordered phase below $15.6\mathrm{K}$, an antiferromagnetic phase below $10\mathrm{K}$ and a low temperature ferrimagnetic phase below $6.5\mathrm{K}$. We identify the low-temperature ferrimagnetic phase from the magnetization and magnetostriction measurements. A ferrimagnetic phase requires two magnetic ions in the unit cell with different magnetic moments, which appears inconsistent with the reported room temperature crystal structure of \ce{CeSb2}. Possible explanations for this are a low temperature structural phase transition, or a charge density wave order. Topographic STM images of the two surface terminations that we observe exhibit an additional superperiodicity similar to the CDW order reported in the La-doped compound \cite{SuderowCDW}, which may render neighbouring magnetic ions inequivalent. 

Quasiparticle interference imaging on termination T2 (CeSb) were conducted above and below the magnetic transitions. These measurements, at high temperature, reveal a dispersing feature which becomes heavier and  a gap opens at $6\mathrm{mV}$ about the Fermi level as the sample enters the lowest temperature magnetic state when cooled to $2.2\mathrm{K}$. This behaviour resembles the opening of a hybridization gap seen in other heavy fermion compounds\cite{Schmidt2010}, but is here linked with entering a magnetically ordered phase. Furthermore a hole-like dispersion is seen in the paramagnetic phase with a maximum at $24.5\mathrm{mV}$ above the Fermi level which shifts by $\sim8\mathrm{mV}$ to lower energy on entering the ferrimagnetic phase (Ferri).  

\section{Conclusion}
In conclusion, from a combination of measurements of the magnetisation as well as of the magnetostriction, we identify the low-temperature phase of \ce{CeSb2} as a ferrimagnetic phase. Based on these measurements, we propose a revised phase diagram of \ce{CeSb2}. STM imaging shows a charge order at the surface with a similar wave vector as the one observed in the charge density wave compound La$_x$Ce$_{1-x}$Sb$_2$ \cite{SuderowCDW}. Spectroscopic measurements of the surface reveal a reconstructed Fermi surface upon entering the low temperature phase. By measuring differential conductance spectroscopy as a function of applied field and temperature we are able to identify a metamagnetic transition at the surface. Measuring the sample's magnetostriction with the STM we are directly able to measure the bulk phase diagram with what is usually considered a surface probe. This demonstrates that STM can be used to measure both surface and bulk magnetic properties simultaneously and opens an intriguing new avenue to magnetic measurement techniques.  

\begin{acknowledgments}
We acknowledge discussions with Andreas Rost and Astrid Schneidewind. CT and PW acknowledge support from EPSRC through EP/R031924/1 and EP/T031441/1. Work at the Ames Laboratory (CA, SLB, PCC) was supported by the U.S. Department of Energy, Office of Science, Basic Energy Sciences, Materials Sciences and Engineering Division. The Ames Laboratory is
operated for the U.S. Department of Energy by Iowa State University under Contract No. DE-AC02-07CH11358.
\end{acknowledgments}

\section*{Appendix}
\renewcommand{\thesection}{Appendix \Alph{section}}
\setcounter{section}{0}

\begin{figure*}
\begin{center}
\includegraphics{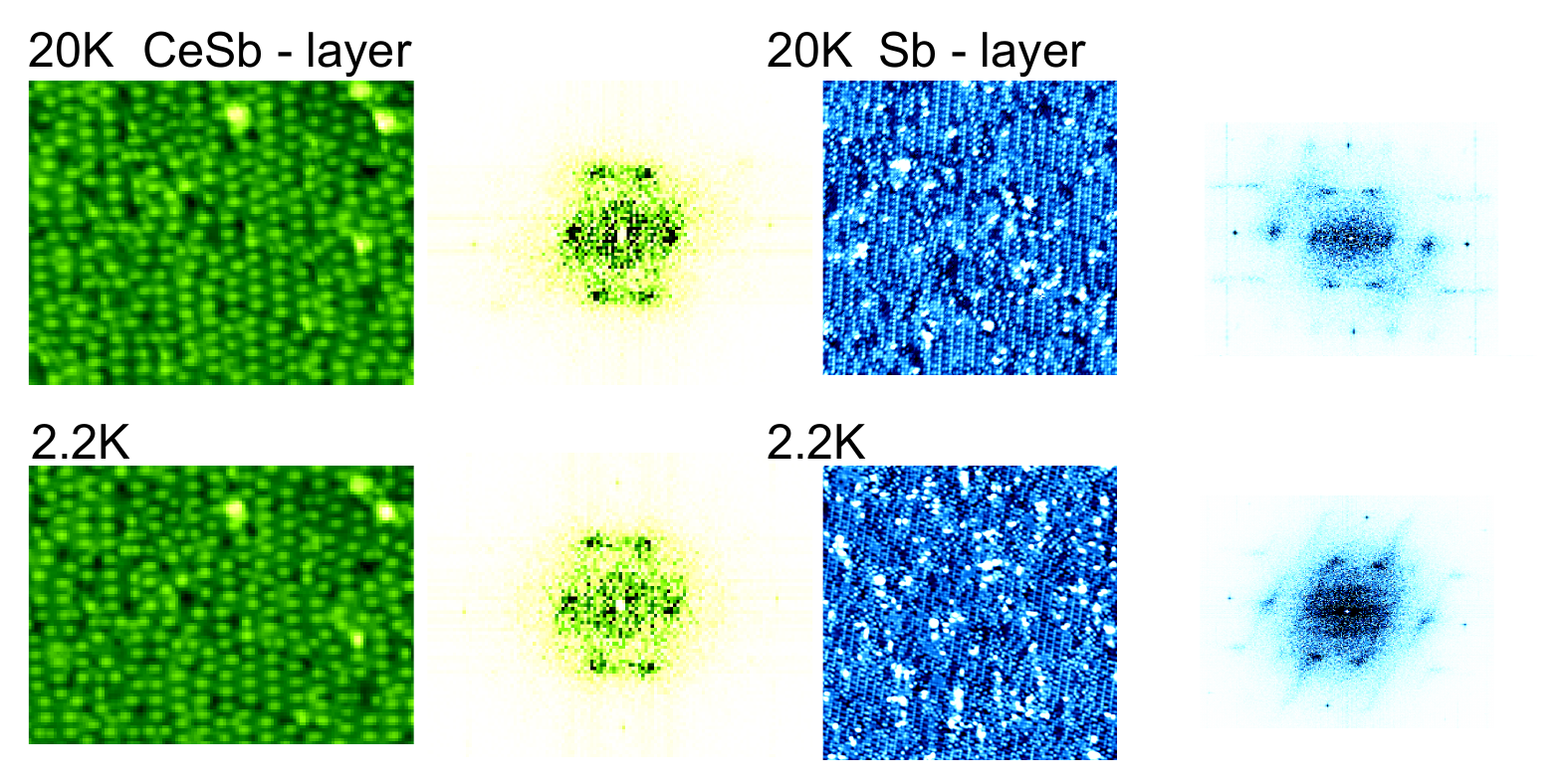}
\end{center}
\caption{Left (green) - Topographies of the termination T2 identified as the CeSb layer recorded at $20\mathrm{K}$ and $2.2\mathrm{K}$ and their respective Fourier transforms ($V=100\mathrm{mV}$, $I=40\mathrm{pA}$). Right (blue) - Topographies of the termination T1 identified as the Sb layer recorded at $20\mathrm{K}$ ($V=300\mathrm{mV}$, $I=50\mathrm{pA}$) and $2.2\mathrm{K}$ ($V=300\mathrm{mV}$, $I=35\mathrm{pA}$) and their respective Fourier transforms.
The same surface features are seen indicating that they are not magnetic in origin.
}
\label{s1}
\end{figure*}

\subsection{Temperature dependent topographic imaging}

Figure~\ref{s1} shows topographic images of the different terminations that have been recorded at temperatures above and below the magnetic phase transitions. None of the observed topographic features are affected by the magnetic phase transition indicating that they are not related to the magnetic order in this compound. Possible origins include a surface reconstruction, a surface charge density wave, or a bulk charge density wave. The fact that the same periodicity is observed on both terminations suggests that the origin is bulk-like.

\begin{figure}[bt!]
\begin{center}
\includegraphics[width=0.8\columnwidth]{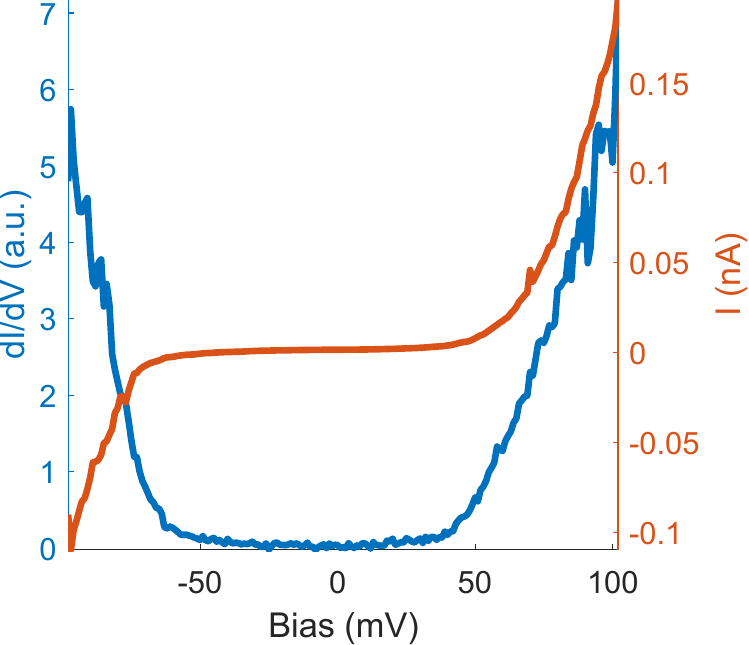}
\end{center}
\caption{A typical differential conductance spectrum $dI/dV(V)$ recorded after the tip has collected material from the termination labelled T1 in the main text (blue line). The corresponding bias-dependence of the tunneling current $I(V)$ is shown in red ($V_\mathrm{set}=100\mathrm{mV}$, $I_\mathrm{set}=200\mathrm{pA}$, $V_{mod}=1\mathrm{mV}$). 
}
\label{s2}
\end{figure}

\subsection{Identification of T1 as Sb termination}
\label{appendix-t1}
Material collected from the termination labelled T1 in the main text yields semiconducting tips which show a gap of a similar magnitude as that of bulk Sb \cite{Huntley_1972} indicating that the material collected from this termination is primarily Sb. Figure~\ref{s2} shows a tunneling spectrum of the semiconducting band gap of a tip prepared on this termination. Tip preparation on T2 typically results in metallic tips.

\begin{figure}[bt!]
\begin{center}
\includegraphics{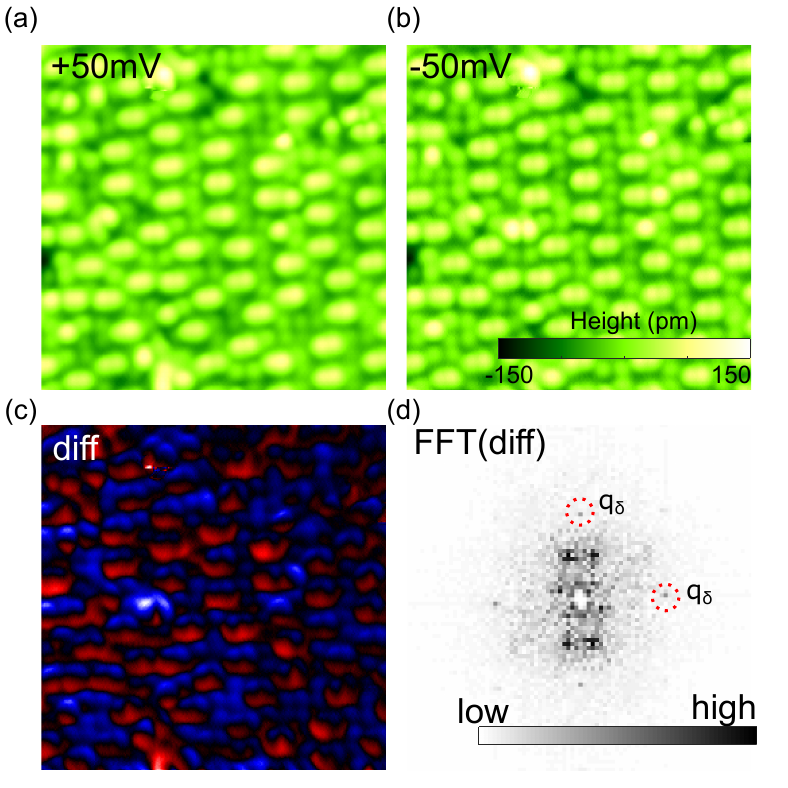}
\end{center}
\caption{(a) Topographic image of the CeSb termination (T2) recorded at positive bias voltage, $V_{\mathrm{S}}=50\mathrm{mV}$, and (b) at negative bias voltage, $V_{\mathrm{S}}=-50\mathrm{mV}$ ($(10 \times 10)\mathrm{nm}^{2}$,  $I_{\mathrm{s}}=40\mathrm{pA}$). (c) The difference between the images shown in (a) and (b), and (d) its Fourier transform.}
\label{s10}
\end{figure}
\subsection{Bias dependent imaging of termination T2}
\label{appendix-t2}
Figure~\ref{s10} shows bias dependent images of the termination T2, which we identify as the CeSb termination. A phase shift of the bubble-like features can be seen in the difference image (figure~\ref{s10}c) between the topographies recorded at $\pm 50\mathrm{mV}$. The ordering corresponds to the vectors observed at $(\pm \frac{1}{12},\frac{1}{2})$ (in units of $q_\delta$) in the Fourier transformation of the topographic images.

\begin{figure}
\begin{center}
\includegraphics[width=0.8\columnwidth]{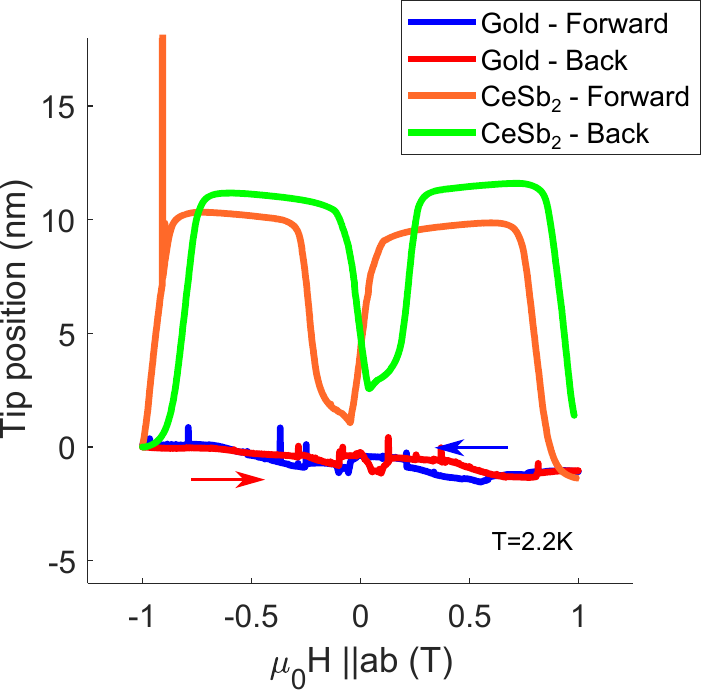}
\end{center}
\caption{Magnetostriction data obtained at $2.2\mathrm{K}$ on a CeSb$_2$ sample and the same measurement recorded on a gold sample for reference. 
}
\label{s5}
\end{figure}

\subsection{Supplementary magnetostriction data}
Figure~\ref{s5} shows a comparison of magnetostriction data obtained by STM on the CeSb$_2$ sample and a gold field emission target used for tip preparation. The size of the response obtained on the CeSb$_2$ sample is significantly larger than that obtained on a non-magnetic gold sample, ruling out any substantial contribution to the magnetostriction from the magnetic response of the internal STM components and sample holder.
Values of the sample magnetostriction are obtained by dividing the change in tip height $\Delta L$ by the sample height $L_0$ along the $c$ axis ($L_0=440\mu\mathrm{m}$). 

\begin{figure*}[bt!]
\begin{center}
\includegraphics{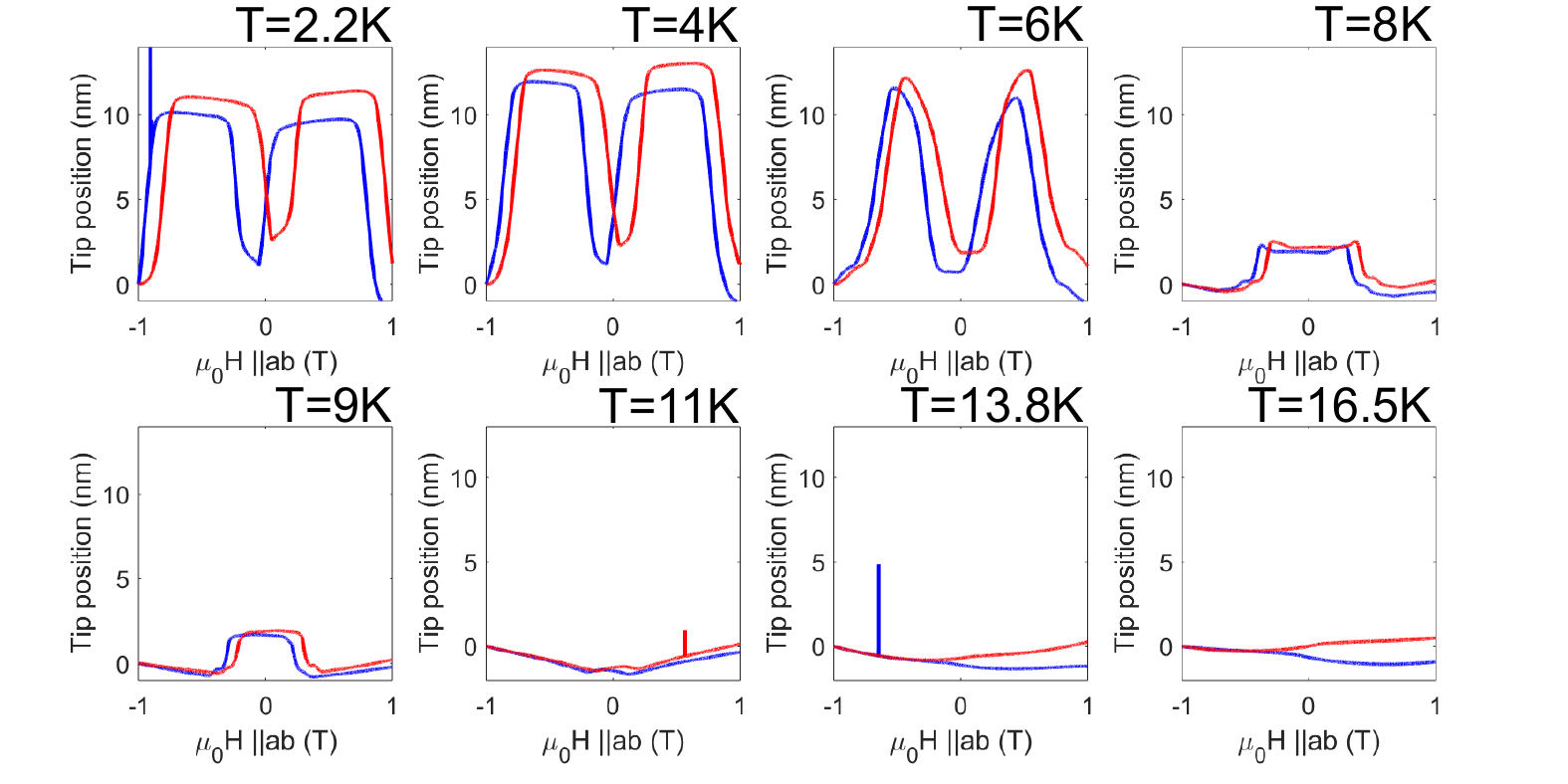}
\end{center}
\caption{STM magnetostriction curves recorded at constant temperatures throughout the magnetic phase diagram of CeSb$_2$. There are clear differences in the magnetostriction in the different phases. The measurements are in the low temperature ferrimagnetic phase below $6.5\mathrm{K}$, antiferromagnetic phase between $6.5\mathrm{K}$ and $\sim11\mathrm{K}$ and the magnetically ordered and paramagnetic phases above $11\mathrm{K}$ and $15.6\mathrm{K}$ respectively.
}
\label{s6}
\end{figure*}

Figure~\ref{s6} shows STM magnetostriction data recorded between positive and negative field of $1\mathrm{T}$ recorded at constant temperatures. Below $6.5\mathrm{K}$, the sample shows a large response to the magnetic field across the low temperature phase. Between $8\mathrm{K}$ and $11\mathrm{K}$, the response is weaker, indicating the onset of the antiferromagnetic phase (AFM). At temperatures above $11\mathrm{K}$, when the sample enters first the magnetic phase (MPh) and then the paramagnetic phase (Para), there is no substantial response of the sample's $c$ axis.  

\begin{figure}[bt!]
\begin{center}
\includegraphics{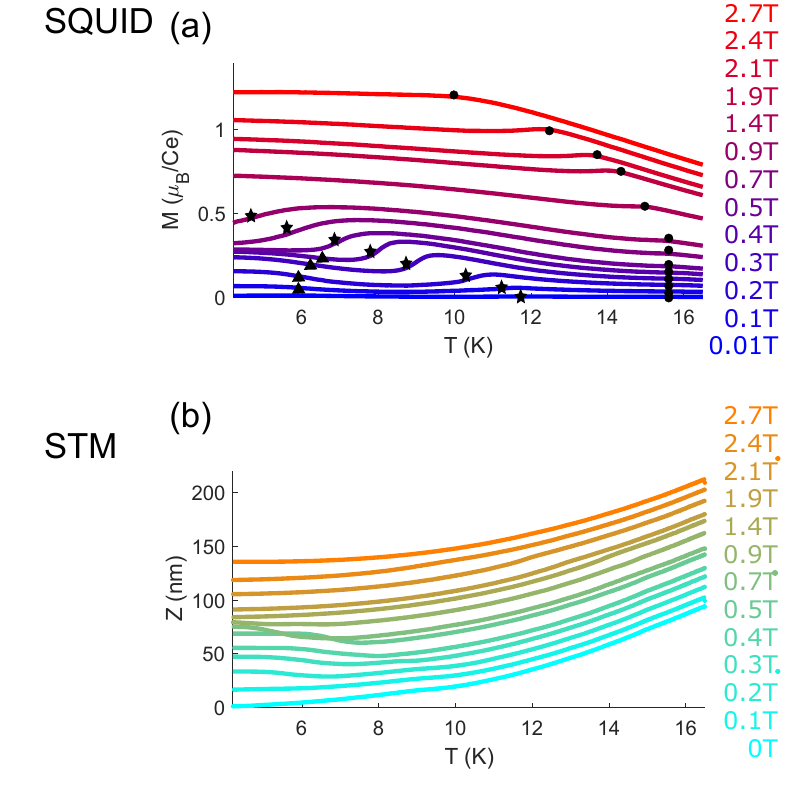}
\end{center}
\caption{(a) Zero field cooled SQUID magnetization measurements $M(H)$ recorded with different applied fields. Derivatives, $\mathrm dM/\mathrm dH$, of which are shown in figure 3 of the main text. Points associated with phase transitions are highlighted. (b) STM tip position versus temperature curves. Curves are vertically offset for clarity. }
\label{s7}
\end{figure}
\subsection{Data used to construct phase diagram}
The magnetization and magnetostriction data used to construct the phase diagram shown in figure 7 of the main text are shown in figure~\ref{s7}. All data were obtained on the same CeSb$_2$ sample. The data in both measurements were measured whilst warming after cooling the sample in zero applied magnetic field. 

\begin{figure}[bt!]
\begin{center}
\includegraphics{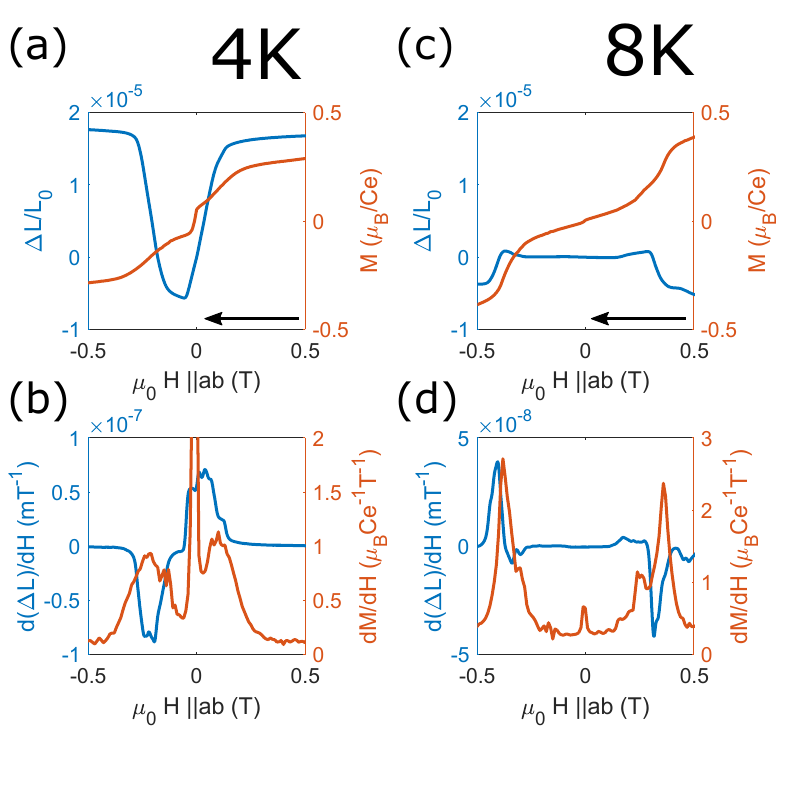}
\caption{(a) Blue curve and axis - Isothermal Magnetostriction data measured in the low temperature ferrimagnetic phase at $4\mathrm{K}$. Orange curve and axis - corresponding magnetization data. Field is swept in the direction of the arrow. (b) The derivatives of the data shown in (a) with respect to the applied field. (c) Blue curve and axis - Isothermal Magnetostriction data measured in the low temperature antiferromagnetic phase at $8\mathrm{K}$. Orange curve and axis - corresponding magnetization data. (d) the derivatives of the data shown in (c) with respect to the applied field.  }
\label{mandm}
\end{center}
\end{figure}

\subsection{Direct comparison of Magnetization and Magnetostriction measurement} 

Figure \ref{mandm} compares both the magnetization and magnetostriction measurement data on the same axes (panels (a) and (c)). The measurements are recorded on the same sample with the field applied in the sample $ab$ plane. The field is orientated at $16^\circ$ from [110] in the magnetostriction data and at a random angle for the magnetization measurement. The derivatives of the data are shown in figure \ref{mandm}(b) and (d). The peaks in the derivatives corresponding to the saturization of the ferrimagnetism in the case of the $4\mathrm{K}$ data and the metamagnetic transition in the case of the $8\mathrm{K}$ data can be seen to occur at very similar fields between the two measurement techniques.

%

\end{document}